# Absorber of Topologically Structured Light

Luka Vignjevic[1], Eric Plum[1], Nikitas Papasimakis[1], and Nikolay I. Zheludev[1,2]

[1]Optoelectronics Research Centre & Centre for Photonic Metamaterials,
University of Southampton, University Road, Southampton SO17 1BJ, UK

[2]Texas A&M University, Institute for Advanced Study, College Station, Texas, 77843-3572, USA

E-mail: l.vignjevic@soton.ac.uk

## Abstract

Polarization- and wavelength-sensitive absorbers for homogeneous electromagnetic waves are crucial in photovoltaics, imaging, and telecommunications. Here, we report on an absorber selective to the topological structure of light. An electromagnetic beam or pulse containing polarization singularities can be fully dissipated by the absorber, while plane waves are rejected regardless of their polarization. The absorber comprises a conical mirror coaxial with the incident propagating beam, which the mirror converts into a standing wave defined by the geometrical Pancharatnam–Berry phase accumulated upon reflection on the mirror. If a "nanowire" absorber is placed along the axis of the cone, singularly-polarized light can create an antinode of the standing wave at the absorber and nearly perfect dissipation of light's energy is achievable regardless of the wavelength. The selective absorber of topologically structured light is of interest for energy harvesting, detection, filtering, and telecommunications applications.

# Introduction

Over the last decade there has been an explosion of interest in the topology of light waves, such as radially and azimuthally polarized light beams and beams carrying orbital angular momentum [1], superoscillatory light [2], and more complex spatiotemporal electromagnetic excitations [3] including space-time nonseparable pulses [4, 5]. The intense interest in topologically structured light is fuelled by possible applications in imaging [6], metrology [7], communication [8, 9], and advanced spectroscopies. To characterize light's topological features, such as phase and polarization singularities, an expanding toolbox has been developed, including interferometric approaches [10-12], mode sorting and decomposition [13], diffraction through gratings and apertures [13-18]. However, said approaches rely on cumbersome setups and/or multiple measurements. Here, we show that a beam or electromagnetic pulse containing a polarization singularity can be absorbed by the proposed device, whereas plane waves of any polarization will be rejected, thus simplifying the process of detecting a polarization singularity to a trivial absorption measurement. As a proof of principle demonstration, we consider radially and azimuthally polarized beams, as well as space-time non-separable toroidal light pulses [19]. We provide a qualitative explanation of the topological light absorber based on geometrical Pancharatnam–Berry phase accumulation and develop its analytical description using Mie scattering theory.

The topological light absorber presented in this paper comprises a conical mirror that shall be coaxial with the propagation direction of the incident light beam and a one-dimensional absorbing element placed along the cone's axis. For practical implementations, such an element can be a nanowire of a diameter much smaller than the wavelength of light. Reflection in the conical mirror converts the incident light into a standing wave centered on the nanowire which allows the regime of either perfect absorption or perfect transmission on the nanowire to be realized, similar to the interference of plane waves on a thin planar absorbing film [20].

Although Mie scattering theory of the absorption process will be considered below, it is insightful to present a toy model of the device based on geometrical optics. We consider linearly polarized light (e.g. in the form of a plane wave or a well-collimated Gaussian beam) propagating collinearly with the cone axis. The response of the absorber is then investigated in two characteristic planes encompassing the axis of the cone: one that is parallel to the polarization of incident light and one that is normal to the incident polarization.

First, we will discuss the counterintuitive vanishing absorption under illumination with a homogeneous plane wave of any polarization. An illustration of this scenario is given in Fig. 1. If we consider a plane containing the incident polarization (see Fig. 1a), light in channels 1 and 2 will undergo polarization rotation at + 90 deg. and - 90 deg, respectively, as indicated by the solid green arrows and shown on the Poincare sphere (Fig. 1b). Depending on the direction of polarization rotation, the geometrical Pancharatnam–Berry phases before and after reflection differ by an amount equal to half of the solid angle encompassed by their trajectory on the Poincare spheres: waves in channels 1 and 2 will gain geometrical phase of $\psi_1 = -\pi/2$ and $\psi_2 = +\pi/2$, respectively. As a result, they will arrive at the nanowire absorber in antiphase creating a node of the standing wave, i.e. zero electric field amplitude at the nanowire. No absorption will take place in the nanowire in this case. In the plane normal to the incident polarization (see Fig. 1c), the polarization of light does not change upon reflection. Waves in channels 3 and 4 arrive at the nanowire with the same phase. However, light with polarization perpendicular to the one-dimensional absorber does not interact with it. No absorption will take place here either. A plane wave of arbitrary polarization can be presented as a superposition of the two scenarios considered here and thus will not be absorbed but it will be rejected by the device.

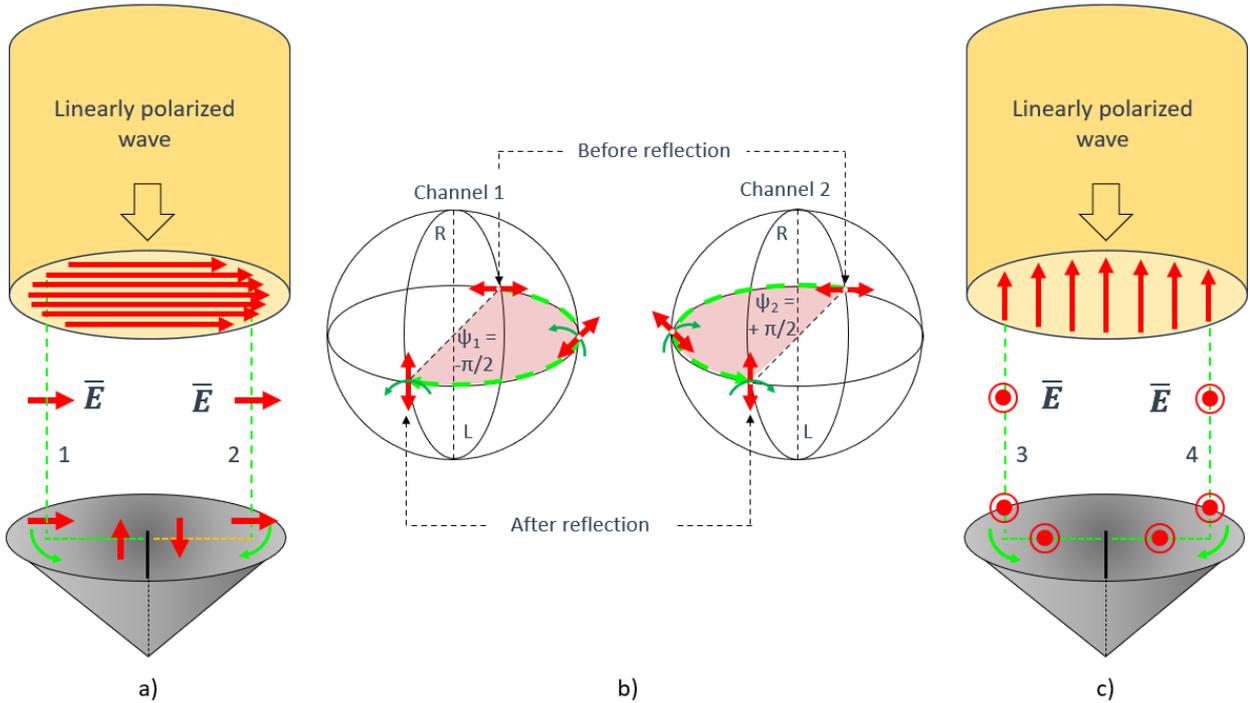

Fig. 1. **Interaction of the topological light absorber with linearly polarized waves.** (a) In the plane that contains the incident polarization (see red arrows) and the axis of the cone, reflection on the conical mirror results in a change of polarization accompanied by accumulation of geometric phase, as illustrated on the Poincare sphere (b). This leads to destructive interference of the reflected fields on the absorber. (c) In the plane that is normal to the incident polarization and contains the axis of the cone, there is no geometric phase accumulation upon reflection, resulting in constructive interference on the one-dimensional absorber. However, the polarization is normal to the absorber axis and as such dissipation is negligible.

Let us now consider two cases of electromagnetic waves containing v point polarization singularities with an index of unity, radially and azimuthally polarized waves, as illustrated in Fig. 2. If the incident wave is radially polarized (see Fig. 2a), light in channels 1 and 2 will undergo polarization rotation at + 90 deg. and - 90 deg., correspondingly, as indicated by the solid green arrows and shown on the Poincare sphere (see Fig. 2b) gaining geometrical phases of $\psi_1 = -\pi/2$ and $\psi_2 = +\pi/2$, correspondingly. Although initially the waves in channels 1 and 2 were in antiphase, they arrive at the nanowire absorber in-phase creating an anti-node of the standing wave and leading to strong absorption of light in the nanowire. On the contrary, if the incident wave is azimuthally polarized, Fig. 2c, the polarization of light does not change upon reflection, and waves in channels 3 and 4 arrive at the nanowire with opposite phases creating a node of the standing wave at the nanowire position. Low absorption will take place in the nanowire in this case. The electric field moduli for the cases of radial and azimuthal polarization are shown in Fig. 2d&e respectively. We can see that the constructive interference, which occurs at the centre of the nanowire for radial polarization, results in a maximum of the electric field at the nanowire. Conversely, the destructive interference, which occurs with azimuthal polarization, results in no electric field at the nanowire centre, thus the device acts as a binary classifier distinguishing between radial and azimuthal polarizations. Additionally, the device is capable of detecting Neel and Bloch type optical skyrmions, since the transverse field structure is the same as that of radially and azimuthally polarized light respectively [21]. It should be noted that while the device can distinguish between radially and azimuthally polarized light, it cannot distinguish between arbitrary beams with V-point singularities, as higher order vectorial beams will be completely rejected or only partially absorbed [22]. The same can be said for C-point singularities, depending on the order and intensity distribution of the beam, they will be completely rejected or partially absorbed [23]. Beams with Mobius strip type polarization structures will be fully rejected as antipode parts of the beam are orthogonally polarized meaning no interference can take place [24]. Light with a phase singularity (orbital angular momentum) is absorbed at 50% as full constructive interference of electric field components along the absorber is allowed only in plane containing the incident polarization (see also Supplementary Section S1).

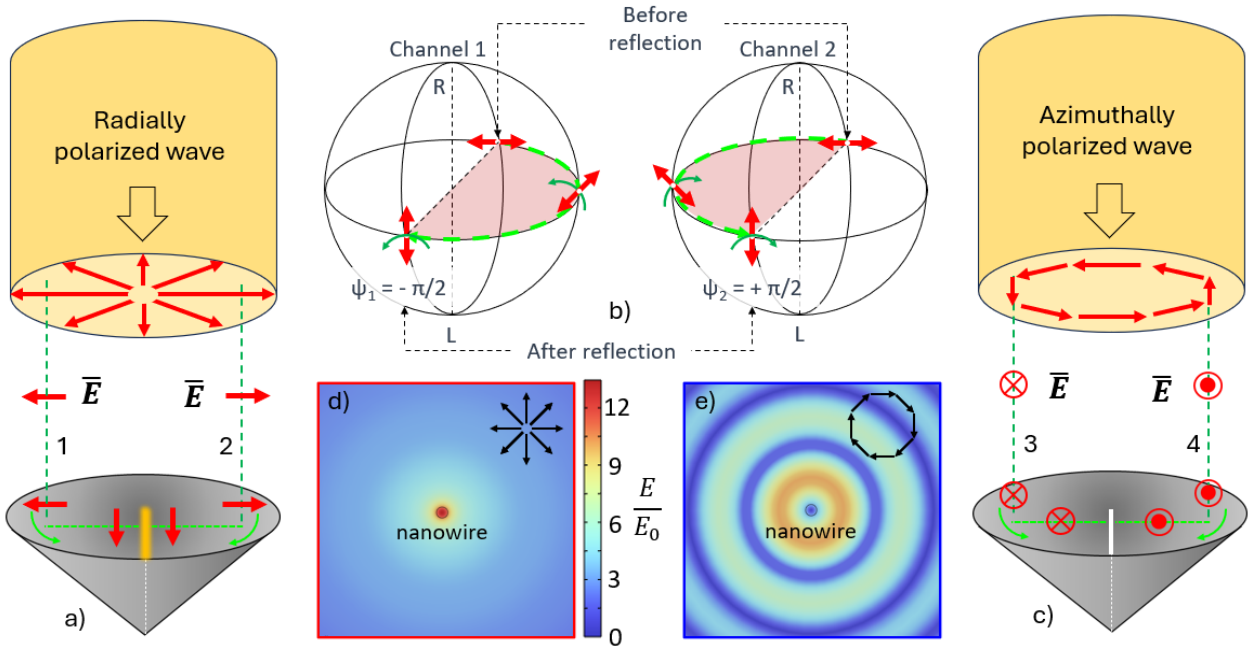

Fig. 2. **Interaction of the light absorber with topologically structured light.** (a) In the case of incident radially polarized light, reflection on the conical mirror results in a change of the polarization state and is accompanied by an accumulation of geometric phase between waves in channels 1 and 2: while initially in antiphase, after reflection the electric field in channels 1 and 2 becomes in-phase leading to constructive interference on the absorber. (b) A pair of Poincare spheres illustrates the change in polarization upon reflection on antipodal positions on the mirror, corresponding to channels 1 and 2. (c) In the case of azimuthally polarized light, there is no geometric phase accumulation upon reflection, resulting in destructive interference on the one-dimensional absorber: the device does not absorb the incident wave. (d-e) Numerically calculated, top-down electric field maps around the nanowire under illumination with radially (d) and azimuthally polarized light (e) showing constructive and destructive interference, respectively. The electric field maps are normalized to the peak amplitude of the incident wave.

The simplified analysis outlined above has demonstrated that light with a polarization singularity can be absorbed by a one-dimensional absorber with a conical illuminator, while homogeneous plane waves of any polarization are rejected.

## Results

Absorption on the nanowire in a cylindrical standing wave is fundamentally similar to the phenomenon of "perfect absorption" and "perfect transmission" on a film of subwavelength thickness in a standing plane wave [25]. In the case of a standing plane wave, the light will be deterministically dissipated if the absorber is placed in the antinode of the wave and exhibits 50% traveling wave absorptivity and 25% reflectivity from either side [26]. To derive conditions for efficient absorption in cylindrical standing wave geometry, we outline an analytical theory of absorption in a one-dimensional absorber with a conical illuminator using the Mie scattering problem [27, 28].

For simplicity, we assume an infinitely long perfectly conducting cone and nanowire of complex refractive index $n=n'+in''$ and radius $\alpha$ that is illuminated by a wave of cylindrical symmetry containing a polarization singularity, with either transverse electric (TE) or transverse magnetic (TM) polarization (e.g. radially or azimuthally polarized light). Here, the amplitude of the corresponding scattered fields is given by (see Supplementary Section S1):

$$b_{TM} = \frac{nJ_0(k\rho)J_0'(nk\rho)|_{\rho=\alpha} - J_0(nk\rho)J_0'(k\rho)|_{\rho=\alpha}}{nH_0^{(1)}(k\rho)J_0'(nk\rho)|_{\rho=\alpha} - J_0(nk\rho)H_0^{(1)'}(k\rho)|_{\rho=\alpha}} \quad (1)$$

$$b_{TE} = \frac{nJ_0(nk\rho)J_0'(k\rho)|_{\rho=\alpha} - J_0(k\rho)J_0'(nk\rho)|_{\rho=\alpha}}{nJ_0(nk\rho)H_0^{(1)'}(k\rho)|_{\rho=\alpha} - H_0^{(1)}(k\rho)J_0'(nk\rho)|_{\rho=\alpha}}, \quad (2)$$

where $k = 2\pi/\lambda$ is the free-space wavevector, $J_0$ and $H_0^{(1)}$ are the 0-th order Bessel and Hankel functions of the first kind, $J_0'$ and $H_0^{(1)'}$ are the corresponding derivatives with respect to the radial coordinate $\rho$. From here absorption in the device can be calculated as:

$$A_{TE} = 1 - 4|0.5 - b_{TE}|^2 \quad (3)$$
$$A_{TM} = 1 - 4|0.5 - b_{TM}|^2 \quad (4)$$

The toy model illustrated in Figs.1&2 assumes infinitely thin absorbing nanowires. The practical realization of the device will inevitably require wire absorbers of finite diameter. This opens opportunities for new functionalities as absorption depends on the refractive index and the radius-to-wavelength ratio, $\alpha/\lambda$. Figures 3a&b present absorption, i.e. the fraction of dissipated energy, as a function of the complex refractive index of the nanowire $n = n' + in''$ for a fixed nanowire radius $\alpha = \lambda/20$ under illumination with radially (a) and azimuthally (b) polarized light. For radially polarized light, a regime close to perfect absorption can be achieved for $n = 3 + 1.1i$. In contrast to infinitely thin absorbers, resonant full absorption can also occur for azimuthally polarized light in the nanowire with a higher refractive index, e.g. $n = 7.5 + 0.2i$ as shown in Fig. 3b. For a finite-thickness nanowire, strong absorption of azimuthally polarized light is possible because the electric field, as described by $J_1(nk\rho)$, reaches maximum inside the nanowire at $nk\rho \cong 1.83$, corresponding to $\rho \cong \lambda/26$. Here $J_1$ is a first order Bessel function of the first kind, $k$ is the free space wavenumber and $\rho$ is the radial position coordinate.

The dependence of absorption on the radius of the nanowire for a fixed real part of the refractive index $n'$ and varying loss $n''$ is shown in Figs. 3c&d. Here, complete absorption for both radially and azimuthally polarized light can occur at $\alpha \cong \lambda/20$ and for $n=3$ and $n=7.5$, respectively. Additionally, complete absorption is observed also at $\alpha \cong \lambda/8$ for azimuthally polarized light. Note, when $\alpha \cong \lambda/20$ and for $n=3$, whilst radially polarised light is completely absorbed, linearly polarised light is only 8% absorbed, which indicates the topologically sensitive behaviour of the device (see also Supplementary Section S3).

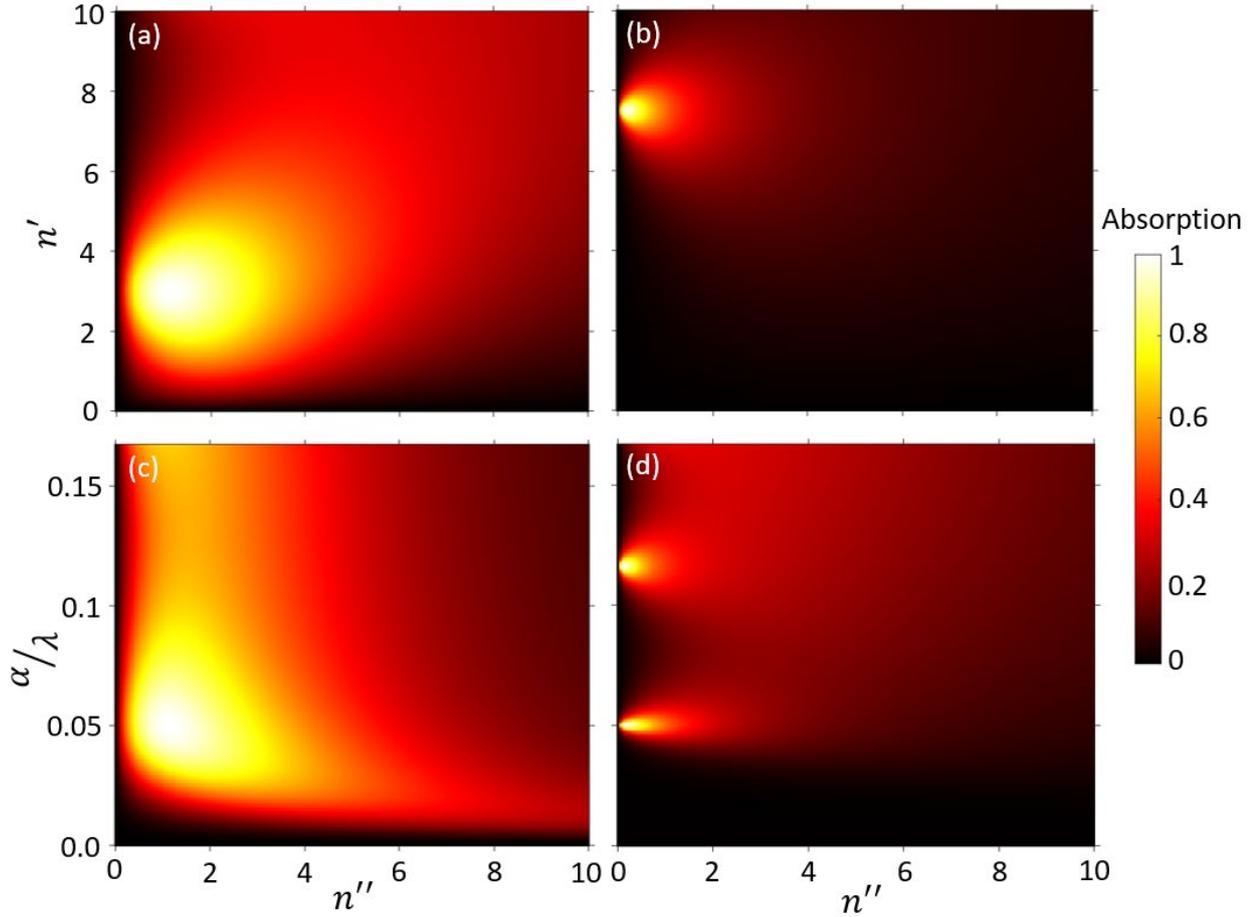

**Fig. 3. Performance of the topological light absorber.** Absorption as a function of the nanowire's complex refractive index under illumination with radially (a) and azimuthally (b) polarized light for a nanowire with radius of $\alpha = \lambda/20$. (c-d) Absorption as a function of nanowire radius and imaginary part of the refractive index, under illumination with radially (c) and azimuthally (d) polarized light. The real part of the nanowire refractive index is $n'=3$ and $n'=7.5$ for (c) and (d), respectively. Panels (b, d) indicate that the conditions for perfect absorption can be reached for azimuthally polarized light too, in the case of high refractive index nanowires.

In a standing light wave, total transmission of light and strong absorption takes place if a thin flat absorber film is placed in either its node or antinode. For a film absorbing 50% and reflecting 25% of the energy of a travelling wave, perfect absorption occurs at a standing wave antinode irrespectively to the wavelength of light [20]. The finite thickness of the absorber and the departure of its reflectance and absorption from the ideal values results in the wavelength dependence of the strength of absorption. Similar behaviour is exhibited by the topological light absorber. Depending on the nanowire radius $\alpha$ the regime of perfect absorption can only be achieved for certain values of the complex refractive index $n = n' + in''$, as seen in Figs. 4a&b, where the refractive index required for complete absorption is plotted as a function of $\alpha/\lambda$.

The concept of the topological light absorber can be practically implemented across the electromagnetic spectrum, subject only to fabrication limitations. Indeed, a microwave device can be manufactured by CNC machining, whereas in the THz range a topological light absorber can be fabricated by a combination of laser micromachining and imprint approaches. Further, deposition assisted by focused ion beam (FIB) allows the fabrication of high-aspect ratio nanostructures, enabling absorbers targeting the mid-IR and optical wavelengths.

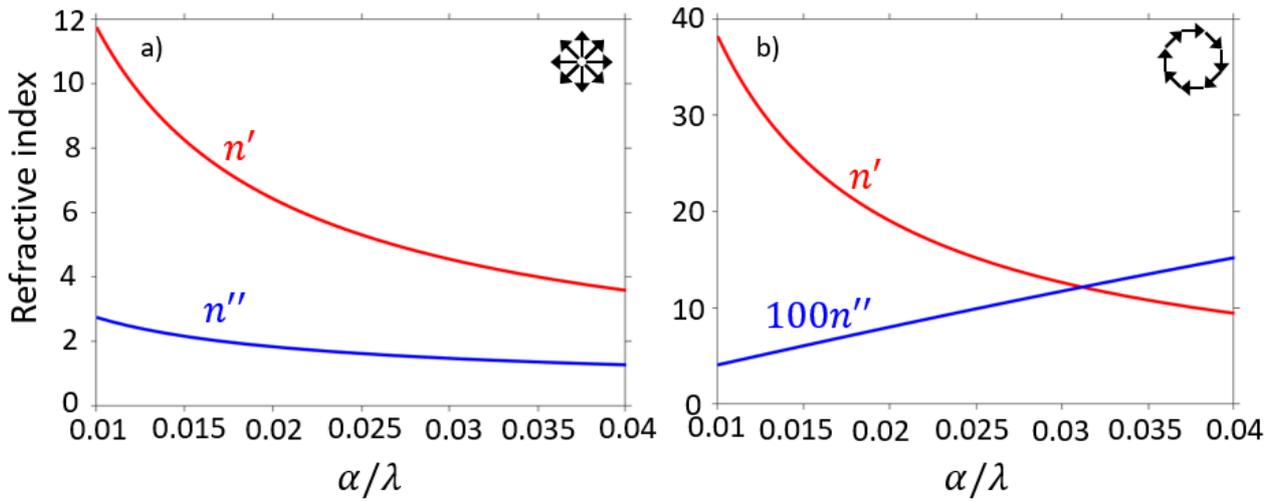

**Fig. 4. Conditions of perfect absorption for nanowires of different radii.** (a) radially polarized light, (b) azimuthally polarized light.

The proposed detector scheme exhibits excellent selectivity for topologically structured electromagnetic pulses. This is illustrated with space-time nonseparable Toroidal Light Pulses that exist in complementary radially (TM) and azimuthally (TE) polarized forms [29]. Using characteristics of a recent experimental demonstration of such pulses [19], i.e. a central wavelength of 800 nm and bandwidth of 200 nm we show that, with a $WTe_2$ nanowire ($\alpha = 32nm$), 98% of the energy of the TM pulse is absorbed, while 98% of the TE pulse is rejected (see Fig. 5). The response of the device to linearly polarized light is similar to that of TE TLPs, they are rejected at levels >90%, which is consistent with the theoretical prediction based on geometric phase outlined in Fig. 1. For similar reasons, circularly polarized pulses will also be rejected (see Supplementary Section S4).

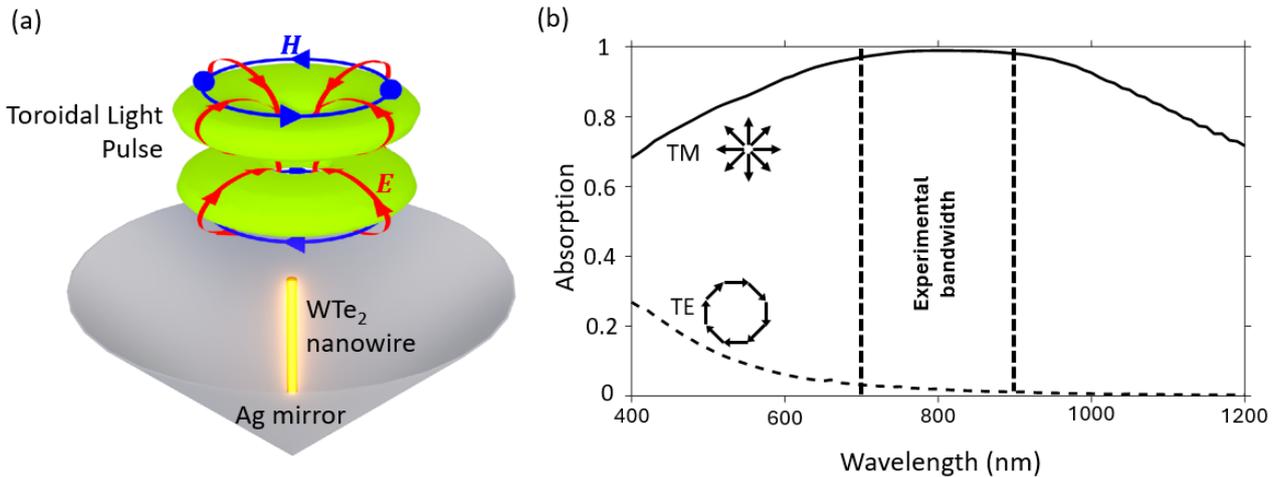

**Fig. 5. Absorption of Toroidal Light Pulses.** (a) Conceptual schematic of topological light absorber and incident TM Toroidal Light Pulses. (b) Spectral dependence of absorption with a $WTe_2$ nanowire. The dashed lines represent the spectral envelope of the experimentally generated Toroidal Light Pulses.

# Conclusion

We report an absorber selective to the topological structure of the incident light, allowing it to be used for topology-selective light management and detection in spectroscopic, telecommunication and metrology applications.

# Acknowledgements

The authors acknowledge the support of the European Research Council (advanced grant FLEET-786851, Funder Id: https://doi.org/10.13039/501100000781) and the Engineering and Physical Sciences Research Council, UK (grant number EP/T02643X/1).

# Supplementary Material

In this section, we present an analytical theory for the description of the topological light absorber. We also discuss its response to topologically trivial light and other forms of topologically structured light.

# Data availability

The data from this paper can be obtained from the University of Southampton ePrints research repository https://doi.org/10.5258/SOTON/DXXX.

# References

1. Shen, Y., et al., *Optical vortices 30 years on: OAM manipulation from topological charge to multiple singularities.* Light: Science & Applications, 2019. **8**(1): p. 90.
2. Zheludev, N.I. and G. Yuan, *Optical superoscillation technologies beyond the diffraction limit.* Nature Reviews Physics, 2022. **4**(1): p. 16-32.
3. Chong, A., et al., *Generation of spatiotemporal optical vortices with controllable transverse orbital angular momentum.* Nature Photonics, 2020. **14**(6): p. 350-354.
4. Murat, Y., et al. *Demonstration of 3D space-time wave packets*. in *Proc.SPIE*. 2022.
5. Shen, Y., N. Papasimakis, and N.I. Zheludev, *Nondiffracting supertoroidal pulses and optical "Kármán vortex streets".* Nature Communications, 2024. **15**(1): p. 4863.
6. Pu, T., et al., *Unlabeled Far-Field Deeply Subwavelength Topological Microscopy (DSTM).* Advanced Science, 2021. **8**(1): p. 2002886.
7. Liu, T., et al., *Picophotonic localization metrology beyond thermal fluctuations.* Nature Materials, 2023. **22**(7): p. 844-847.
8. Zhu, Z., et al., *Compensation-free high-dimensional free-space optical communication using turbulence-resilient vector beams.* Nature Communications, 2021. **12**(1): p. 1666.
9. Nape, I., et al., *Revealing the invariance of vectorial structured light in complex media.* Nature Photonics, 2022. **16**(7): p. 538-546.
10. Kulkarni, G., et al., *Single-shot measurement of the orbital-angular-momentum spectrum of light.* Nature Communications, 2017. **8**(1): p. 1054.
11. Soskin, M.S., et al., *Topological charge and angular momentum of light beams carrying optical vortices.* Physical Review A, 1997. **56**(5): p. 4064-4075.
12. Angelsky, O.V., et al., *Interferometric methods in diagnostics of polarization singularities.* Physical Review E, 2002. **65**(3): p. 036602.

13. Forbes, A., A. Dudley, and M. McLaren, *Creation and detection of optical modes with spatial light modulators.* Advances in Optics and Photonics, 2016. **8**(2): p. 200-227.
14. Moreno, I., et al., *Decomposition of radially and azimuthally polarized beams using a circular-polarization and vortex-sensing diffraction grating.* Optics Express, 2010. **18**(7): p. 7173-7183.
15. Hickmann, J.M., et al., *Unveiling a Truncated Optical Lattice Associated with a Triangular Aperture Using Light's Orbital Angular Momentum.* Physical Review Letters, 2010. **105**(5): p. 053904.
16. Zheng, S. and J. Wang, *Measuring Orbital Angular Momentum (OAM) States of Vortex Beams with Annular Gratings.* Scientific Reports, 2017. **7**(1): p. 40781.
17. Feng, F., et al., *On-chip plasmonic spin-Hall nanograting for simultaneously detecting phase and polarization singularities.* Light: Science & Applications, 2020. **9**(1): p. 95.
18. Guo, Y., et al., *Spin-decoupled metasurface for simultaneous detection of spin and orbital angular momenta via momentum transformation.* Light: Science & Applications, 2021. **10**(1): p. 63.
19. Zdagkas, A., et al., *Observation of toroidal pulses of light.* Nature Photonics, 2022. **16**(7): p. 523-528.
20. Zhang, J., K.F. MacDonald, and N.I. Zheludev, *Controlling light-with-light without nonlinearity.* Light: Science & Applications, 2012. **1**(7): p. e18-e18.
21. Shen, Y., et al., *Optical skyrmions and other topological quasiparticles of light.* Nature Photonics, 2024. **18**(1): p. 15-25.
22. Vyas, S., Y. Kozawa, and S. Sato, *Polarization singularities in superposition of vector beams.* Optics Express, 2013. **21**(7): p. 8972-8986.
23. Senthilkumaran, P., *Polarization singularities*, in *Singularities in Physics and Engineering*. 2018, IOP Publishing. p. 9-1-9-46.
24. Galvez, E.J., et al., *Multitwist Möbius Strips and Twisted Ribbons in the Polarization of Paraxial Light Beams.* Scientific Reports, 2017. **7**(1): p. 13653.
25. Chong, Y.D., et al., *Coherent Perfect Absorbers: Time-Reversed Lasers.* Physical Review Letters, 2010. **105**(5): p. 053901.
26. Roger, T., et al., *Coherent perfect absorption in deeply subwavelength films in the single-photon regime.* Nature Communications, 2015. **6**(1): p. 7031.
27. *Light scattering by small particles. By H. C. van de Hulst. New York (John Wiley and Sons), London (Chapman and Hall), 1957. Pp. xiii, 470; 103 Figs.; 46 Tables. 96s.* Quarterly Journal of the Royal Meteorological Society, 1958. **84**(360): p. 198-199.
28. Bohren, C.F. and D.R. Huffman, *A Potpourri of Particles*, in *Absorption and Scattering of Light by Small Particles*. 1998. p. 181-223.
29. Hellwarth, R.W. and P. Nouchi, *Focused one-cycle electromagnetic pulses.* Physical Review E, 1996. **54**(1): p. 889-895.